# Control of spin-orbit torque-driven domain nucleation through geometry in chirally coupled magnetic tracks

Guillaume Beaulieu[1,2], Zhaochu Luo[1,2,3], Víctor Raposo[4], Laura J. Heyderman[1, 2], Pietro Gambardella[5], Eduardo Martínez[4], Aleš Hrabec[1,2,*]

[1]Laboratory for Mesoscopic Systems, Department of Materials, ETH Zurich, 8093 Zurich, Switzerland

[2]Laboratory for Multiscale Materials Experiments, PSI Center for Neutron and Muon Sciences, Forschungsstrasse 111, 5232 Villigen PSI, Switzerland

[3]State Key Laboratory of Artificial Microstructure and Mesoscopic Physics, School of Physics, Peking University, 100871 Beijing, China

[4]Universidad de Salamanca, Department of Applied Physics, E-37008 Salamanca, Spain

[5]Laboratory for Magnetism and Interface Physics, Department of Materials, ETH Zurich, 8093 Zurich, Switzerland

[*]Author to whom correspondence should be addressed: ales.hrabec@psi.ch

**The interfacial Dzyaloshinskii-Moriya interaction (DMI) can be exploited in magnetic thin films to realize lateral chirally coupled systems, providing a way to couple different sections of a magnetic racetrack and realize interconnected networks of magnetic logic gates. Here, we systematically investigate the interplay between spin-orbit torques, chiral coupling and the device design in domain wall racetracks. We show that the current-induced domain nucleation process can be tuned between single-domain nucleation and repeated nucleation of alternate domains by changing the orientation of an in-plane patterned magnetic region within an out-of-plane magnetic racetrack. Furthermore, by combining experiments and micromagnetic simulations, we show that the combination of damping-like and field-like spin-orbit torques with DMI results in selective domain wall injection in one of two arms of a Y-shaped devices depending on the current density. Such an element constitutes the basis of domain wall based demultiplexer, which is essential for distributing a single input to any one of the multiple outputs in logic circuits. Our results provide input for the design of reliable and multifunctional domain-wall circuits based on chirally coupled interfaces.**

In heterostructures comprising heavy metal and magnetic thin films, the presence of structural inversion asymmetry and strong spin-orbit coupling can lead to a net Dzyaloshinskii-Moriya interaction (DMI) [1–3]. The DMI favors an orthogonal alignment of adjacent magnetic moments [4] and is known to stabilize the Néel-type domain walls (DWs) with a fixed chirality in magnetic thin films with perpendicular magnetic anisotropy [5–8] as shown schematically in the left-hand panels of Fig. 1(a). The dynamics of chiral Néel DWs driven by spin-orbit torques (SOTs), arising from the spin Hall and Rashba-Edelstein effects [9], has led to the emergence of efficient and fast current-driven DW motion [10–16]. The promise of high density, as well as performant and enduring memory devices relying on current-driven DW motion has motivated a significant amount of theoretical and experimental studies [17]. The SOTs determining the DW dynamics are the damping-like torque $\boldsymbol{T}^{\mathrm{DL}} = T_{\mathrm{DL}}\boldsymbol{m} \times (\boldsymbol{\sigma} \times \boldsymbol{m})$ and the field-like torque $\boldsymbol{T}^{\mathrm{FL}} = T_{\mathrm{FL}}\boldsymbol{m} \times \boldsymbol{\sigma}$, where $\boldsymbol{\sigma}$ is the spin accumulation at the heavy metal/ferromagnet interface. The magnitude of both types of torques generally depends on the parameters of the multilayer such as the choice of materials or their thickness [18,19]. The damping-like torque usually dominates and its two components contribute to the DW motion and the tilt [20–22], as illustrated in the right-hand panels of Fig. 1(a). The in-plane part of the damping-like torque causes a rotation of the magnetic moments in the center of the DW towards the spin accumulation direction and, in response to this, the DW itself develops a tilt in order to preserve the chirality imposed by the DMI. The DW tilt has a disruptive effect on the performance of the memory devices, causing the DWs to move at different speeds, and a solution for this issue was proposed using multilayers based on synthetic antiferromagnets [23]. The field-like torque is generally neglected as it plays a less important role in magnetization dynamics [24]. However, it can lower the critical current densities required for magnetization switching [25–28] and affect the domain wall tilt [14,20,21] as well as promote domain nucleation at specific sites [14].

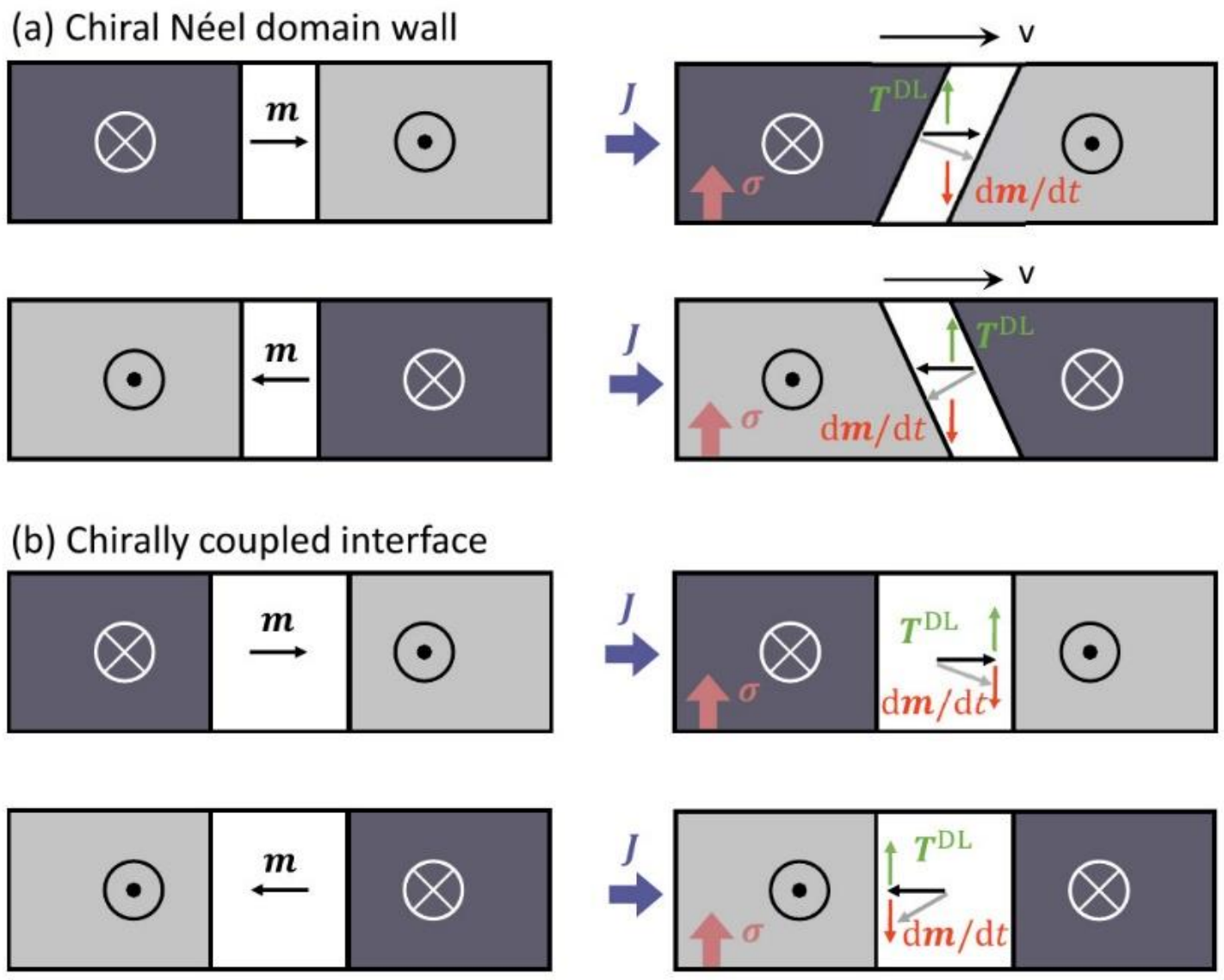

**Figure 1. Principle of SOT action on IP magnetized chiral textures.** *(a) Top-view schematics of the chiral DW in the ⊗⊙ (upper panel) and ⊙⊗ (lower panel) configuration, assuming left-handed Néel walls. When an electric current* $\boldsymbol{J}$ *is applied (purple), a spin accumulation* $\boldsymbol{\sigma}$ *at the interface exerts predominantly the DL torque* $\boldsymbol{T}^{DL}$ *(green). In response to this, the DW itself develops a tilt. (b) Top-view schematics of the chirally coupled magnetic texture in an OOP-IP-OOP environment. When electric current is applied, the magnetization inside the IP regions is tilted away from the DMI-enforced left-handed configuration.*

More recently, the interfacial DMI and SOT-driven DW motion have been integrated in nanomagnetic systems with hybrid magnetic anisotropy. By locally tuning the magnetic anisotropy of a racetrack from an in-plane (IP) to an out-of-plane (OOP) configuration in a multilayer with sizeable DMI, it was shown that the chiral states depicted in Fig. 1(b) are strongly favored [29–31]. This is because the energy is lowered by an amount proportional to $\boldsymbol{D} \cdot (\boldsymbol{m_1} \times \boldsymbol{m_2})$, where the DMI strength is expressed by *D*, and $\boldsymbol{m}_1$ and $\boldsymbol{m}_2$ represent two neighboring magnetic moments. This provided a way to develop all-electric DW-based logic architectures [32], a current-driven inverter DW injector [33], a DW diode [34], reconfigurable DW logic gates [35] and chiral vortex oscillators [36]. In these structures, SOTs were used to drive DWs in the OOP regions and across the IP regions. Nevertheless, one can expect that the SOTs also to act on the IP magnetized region incorporated in the OOP racetrack, as depicted in Fig. 1(b), where the damping-like torque predominantly causes a rotation of the IP magnetization away from the orientation imposed by DMI. Unlike DWs, which are able to tilt in order to preserve the chirality, the SOTs in the chirally coupled interface force the IP magnetic moment to violate the DMI favoured state.

In this work, we show that the current-induced magnetization tilt in the IP region has a key impact on the nucleation process of a reverse domain in the OOP region. We investigate how SOTs couple to chiral orthogonal (OOP-IP) magnetic anisotropy regions and demonstrate how the interplay of these effects can be tuned to give different forms of domain nucleation, ranging from a single-domain nucleation to a repeated nucleation of alternate domains, depending on the orientation of the in-plane patterned magnetic region, the initial magnetic orientation and current amplitude. Our findings also provide a deeper insight into the role of the damping-like and field-like torques in determining the switching mechanism in chirally coupled magnets, which are important to fully exploit spin-orbit related effects in magnetic racetracks.

The magnetic racetracks were patterned from a Pt (6 nm)/Co (1.6 nm)/Al (2 nm) trilayer film using electron beam lithography combined with lift off processing. The trilayer films were deposited using dc magnetron sputtering at a base pressure of $<2 \times 10^{-8}$ Torr and a deposition Ar pressure of 3 mTorr on Si/$SiN_x$ (200 nm) substrates. The as-grown Co films have an IP magnetic anisotropy which is then tuned in specific regions to be OOP by a selective oxidation of the Al top layer [29]. The regions with IP and OOP anisotropy are thus defined with a high-resolution polymethyl methacrylate (PMMA) positive-resist mask on top of the Al layer patterned by electron-beam lithography. The controlled oxidation process is carried out by exposing the unprotected Al regions with a low-power (30 W) oxygen plasma at an oxygen pressure of 10 mTorr. The devices are then created by milling of the upper Co/Al bilayer through a second PMMA mask patterned by electron beam lithography. The remaining Pt underlayer forms a uniform electric conductive channel to drive the DWs. Finally, Cr (5 nm)/Au (50 nm) electrodes are deposited by electron-beam evaporation and patterned using a further electron-beam lift-off lithography process. With this procedure it is possible to gain a precise control of the lateral chiral coupling by combining the large DMI arising at the Pt/Co interface and the tunable magnetic anisotropy. The oxidation determines the strength of the perpendicular anisotropy, which is maximal for full oxidation of the Co/Al interface [37].

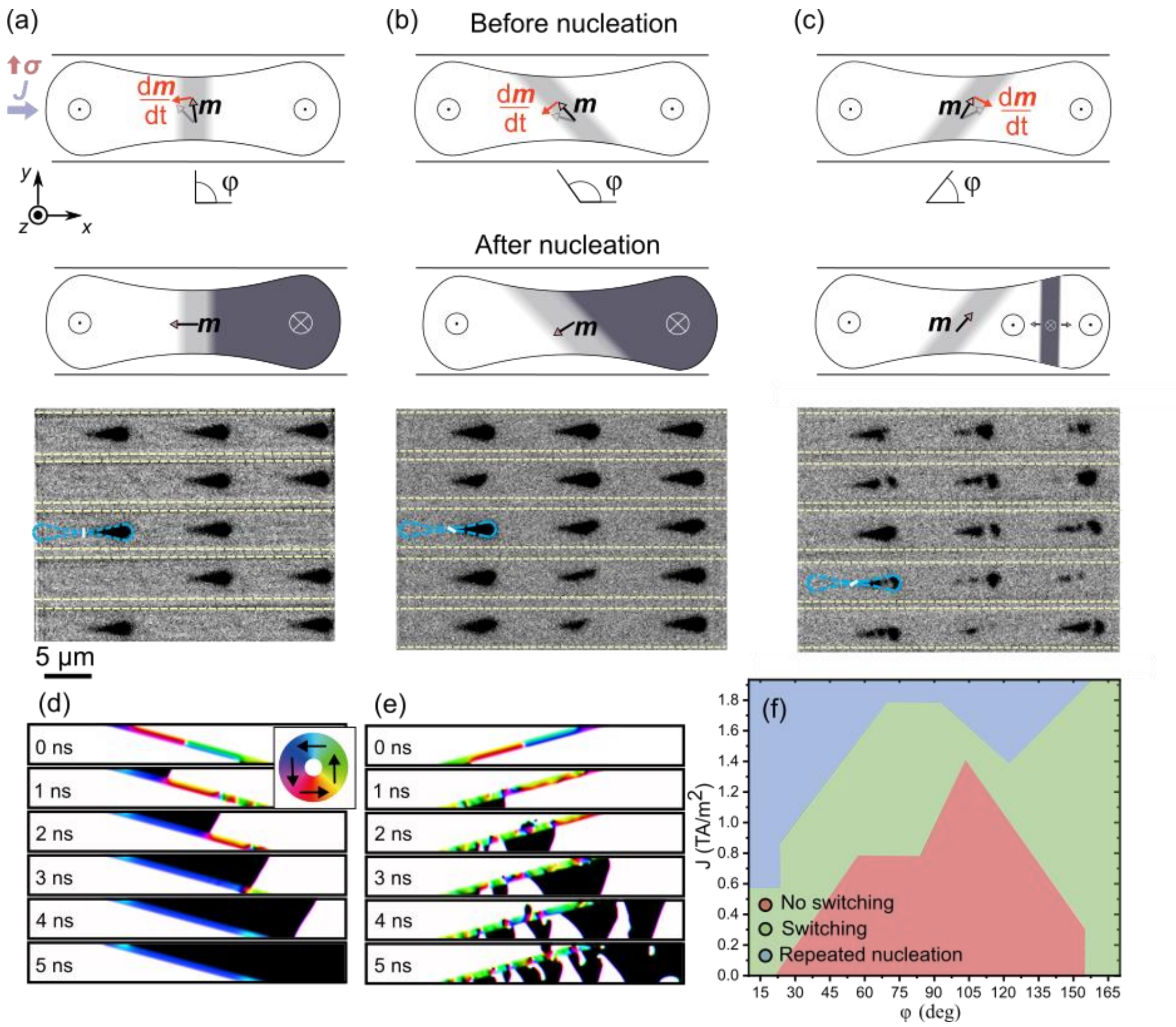

**Fig. 2. Principle of SOT-driven domain nucleation in the chirally coupled region.** *(a-c) Upper row: Top view schematic of the OOP-IP-OOP structure with an IP region (in gray) oriented at (a)* $\varphi = 90°$*, (b)* $\varphi > 90°$ *and (c)* $\varphi < 90°$*. The resulting magnetic configuration is shown in the central row. In the lower row, differential MOKE images show magnetic state after current injection where the edges of the magnetic racetracks are indicated by blue lines, the shape of Pt conduits by yellow lines and the position of the IP region is indicated by a white line. Dark (bright) contrast corresponds to the* ⊗ (⊙) *magnetic state. (d-e) Snapshots of micromagnetic simulations for (d)* $\varphi = 175°$ *and (e)* $\varphi = 15°$*. (f) Phase diagram of the domain switching process as a function of current density J and the angle of the IP region* $\varphi$*. For these simulations, the initial magnetic state was relaxed from an* ⊙ *state in both the OOP and IP anisotropy regions and* $k = \frac{T^{\mathrm{FL}}}{T^{\mathrm{DL}}} = -0.08$*. All experiments and simulations are performed at* $H_z$*=0.*

To characterize the SOT-driven switching process, we have fabricated the racetracks shown in Fig. 2, where the 50 nm-wide IP magnetized region is oriented at various angles $\varphi$ with respect to the current applied along the *x* axis. The DW motion was tracked using a polar magneto-optical Kerr effect (MOKE) microscope. Optimal magnetic contrast was achieved by subtracting a background image of the OOP magnetic saturated racetrack with all subsequent images. For each measurement, the magnetic racetracks were first saturated in the OOP orientation by applying a 200 mT magnetic field, which results in the remanent state that is uniformly magnetized on either side of the IP region. The magnetization in the IP region is then aligned along the long axis of the IP region [see top row of Fig. 2(a-c)] and the magnetic configuration violates the left-handed Néel state imposed by DMI.

In the next step, a dc current with current density of $J = 8 \times 10^{11}$ A/m$^2$ was applied along $x$ in order to nucleate and displace DWs. As can be seen in the Kerr micrograph in Fig. 2(a), for $\varphi = 90°$, only some of the devices switch their magnetic state. Here, the spin accumulation $\boldsymbol{\sigma}$ and magnetization in the center of the DW are collinear, so we would expect that there is no torque exerted. In this case, a torque will only arise from the thermal excitations or deviations in the orientation of the magnetization at the edges of the wire. The switching process can be improved by changing the orientation of the IP region $\varphi$ as shown in Fig. 2(b) and, in this case, we achieve magnetization switching in all 15 devices. As illustrated in Fig. 2(b), here the torque tilts the magnetization towards an ⊙←⊙ state, which can then lower its energy by nucleating a reverse domain, so forming a DMI-compatible ⊙←⊗ state. Such an orientation of the IP region also leads to the creation of a preferred nucleation site in the OOP region at the corner where the IP region meets the edge [34]. In stark contrast, the nucleation process is highly disrupted when the IP region is oriented in the opposite direction. In the case depicted in Fig. 2(c), the SOT tilts the IP magnetization towards $+x$, forming an ⊙→⊙ state. Here, the nucleation of a reverse domain would lead to a DMI-violating (i.e. right-handed) ⊙→⊗ state. Such a competition between DMI, which tries to form a left-handed magnetic texture, and $\boldsymbol{T}^{\mathrm{DL}}$, which favor right-handed chirality, can then lead to the generation of a sequence of domains with alternating ⊙ and ⊗ magnetization. As can be seen from the Kerr micrographs in Fig. 2(c), the consequence of the current application is, indeed, a multidomain state.

Since the processes occurring in the narrow IP region are far below the temporal and spatial resolution of our Kerr microscopy measurements, we turn to micromagnetic modelling using the MuMax3 software [38] to obtain insights into the details of the nucleation process. Here, the cell size was set to $2 \times 2 \times 1.6\ \mathrm{nm}^3$. The magnetic parameters used in the simulations are based on the prototypical Pt/Co/AlOx trilayer film [29], with a saturation magnetization of $M_s = 0.4$ MA/m, an exchange constant of $A = 12$ pJ/m, a uniaxial anisotropy constant of $K_0 = 0.4\ \mathrm{MJ/m^3}$ in the OOP anisotropy region, and $\alpha = 0.1$ for the Gilbert damping parameter. The interfacial DMI strength was set to be $D = 1.2\ \mathrm{mJ/m^2}$, favoring left-handed Néel states, and the spin Hall angle $\theta_{SH} = +0.1$, which means that, if the current $J$ is flowing along $+x$, the resulting spin accumulation $\boldsymbol{\sigma}$ points along $+y$, as indicated in Fig. 1. In specific cases, disorder was included by introducing grain-to-grain variations affecting the distribution in anisotropy, DMI and magnetization [39]. The average grain size is set to 10 nm and the grains are arranged in a Voronoi fashion. The width of the IP region is fixed to be 50 nm. We have considered both damping-like and field-like torques, and their mutual amplitude can be expressed by a coefficient $k$, so that $T^{\mathrm{FL}} = kT^{\mathrm{DL}}$. Prior to each simulation run, the initial state is reached by relaxing the system from a state with OOP magnetization in the OOP and IP anisotropy regions.

The micromagnetic simulations confirm that the final state depends on the orientation $\varphi$ of the IP region. While a single nucleation is observed for $\varphi = 175°$ as shown in Fig. 2(d), a multidomain state results from a repeated domain nucleation process at the IP magnetized region for $\varphi = 15°$ (with $k = -0.08$ [40, 41]) as can be seen in Fig. 2(e). This can be understood by considering the action of the damping-like torque on the IP magnetization. While the torque orients the IP magnetic moment towards the left-handed state favored by the DMI prior to the nucleation for $\varphi > 90°$, as shown schematically in the top panel in Fig. 2(b), it tilts the magnetization towards the right-handed state for $\varphi < 90°$ as shown schematically in the top panel in Fig. 2(c). During the application of the dc current (1 s-long pulse), the SOT-induced tilt of the magnetization and DMI compete, which results in the aforementioned repeated nucleation events. The process is reversed when the magnetization is initialized with opposite OOP magnetic field, i.e. the repeated domain nucleation is observed for $\varphi = 175°$. A phase diagram for the mechanism of the domain nucleation, as a function of the orientation of the IP region $\varphi$ and current density $J$, is given in Fig. 2(f). The strong asymmetry in the diagram between $\varphi < 90°$ and $\varphi > 90°$ is visible, highlighting the importance the orientation of the IP region.

In order to investigate the switching process in more complex structures, which were previously used to construct NAND/NOR logic gates [32], we have fabricated the Y-shaped devices depicted in Fig. 3(a). The racetrack is composed of one upstream and two downstream tracks separated by an IP magnetized V-shaped region. The electric current is applied along the +*x* direction. According to the results of the experiment with the IP magnetized regions oriented at different angles described above, one can expect that one racetrack will generally favour a single nucleation while the other will favor the repeating nucleation process.

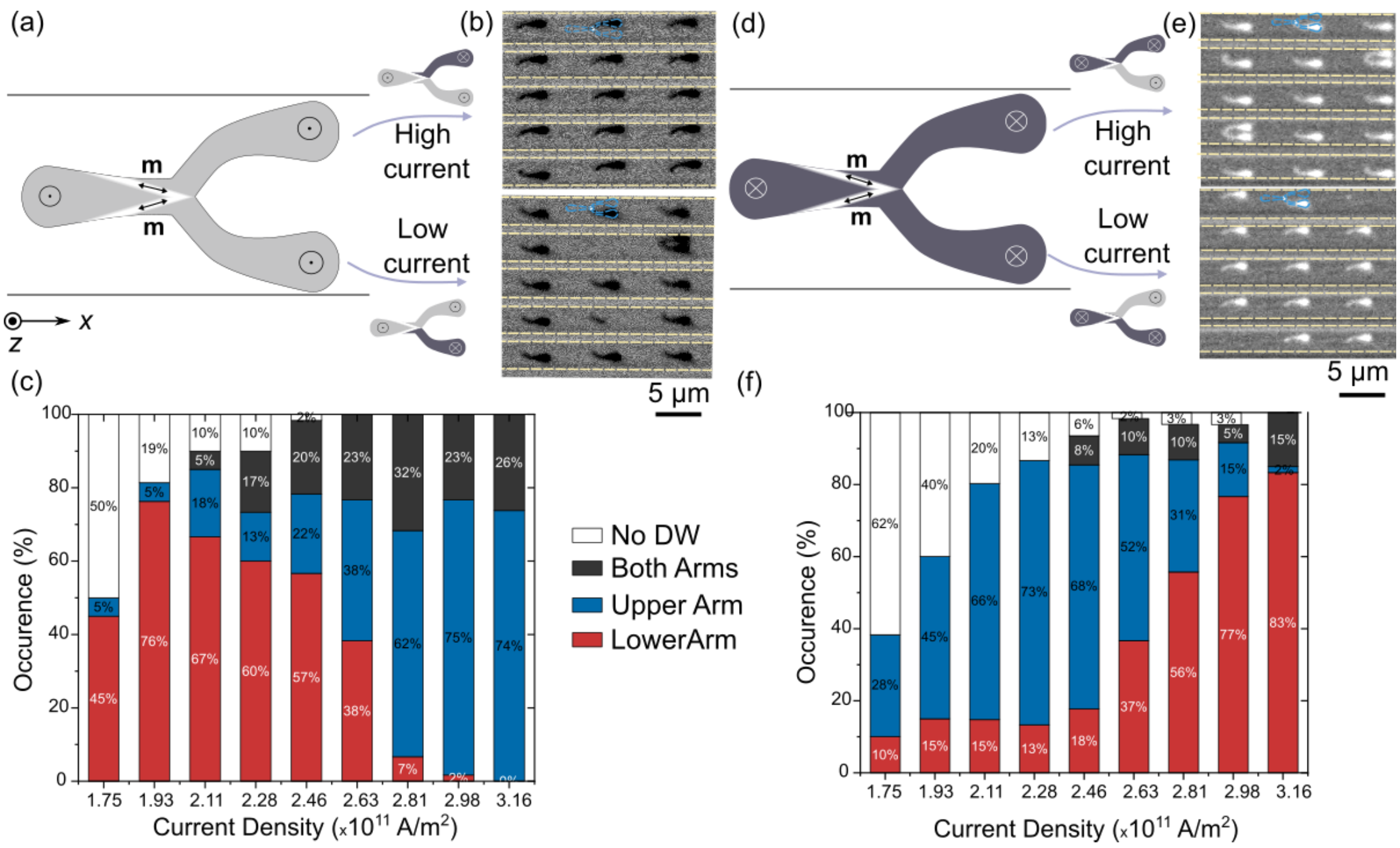

**Fig. 3. Current amplitude-dependent switching in Y-shaped devices.** (a) *Schematic of the Y-shaped device with the magnetic configuration after applying the initialization magnetic field +$H_z$ along the z direction. (b) Snapshot of the resulting Kerr contrast after application of a current with J*=0.23 $TA/m^2$ *(low current) and with* $J = 0.32\ TA/m^2$ *(high current) at $H_z$=0. (c) Resulting magnetic state as a function of current density. (d-f) The same as (a-c) but now for a racetrack initially saturated with a -$H_z$ field.*

The magnetic racetracks are first saturated by applying a magnetic field +$H_z$ and then a dc current (1-s-long pulse) of various amplitude is applied. Surprisingly, at low current densities, only the lower racetrack switches its magnetic state while at high current densities only the upper racetrack switches [Fig. 3(b)]. This selective switching is thus in contrast with our expectations based on the experiments described above, where a $\varphi$ < *90°* orientation in one of the arms supports a single nucleation event while $\varphi$ > *90°* orientation in the other arm would support a repeated generation of reverse domains. We have made a systematic characterization of the switching process as a function of current density, with the results summarized in Fig. 3(c). Here one can see a gradual change in behaviour going from lower to upper arm magnetization switching as the current density is increased. This process is reversed when the sign of the initializing magnetic field is reversed giving -$H_z$, as shown experimentally in Fig. 3(d-f).

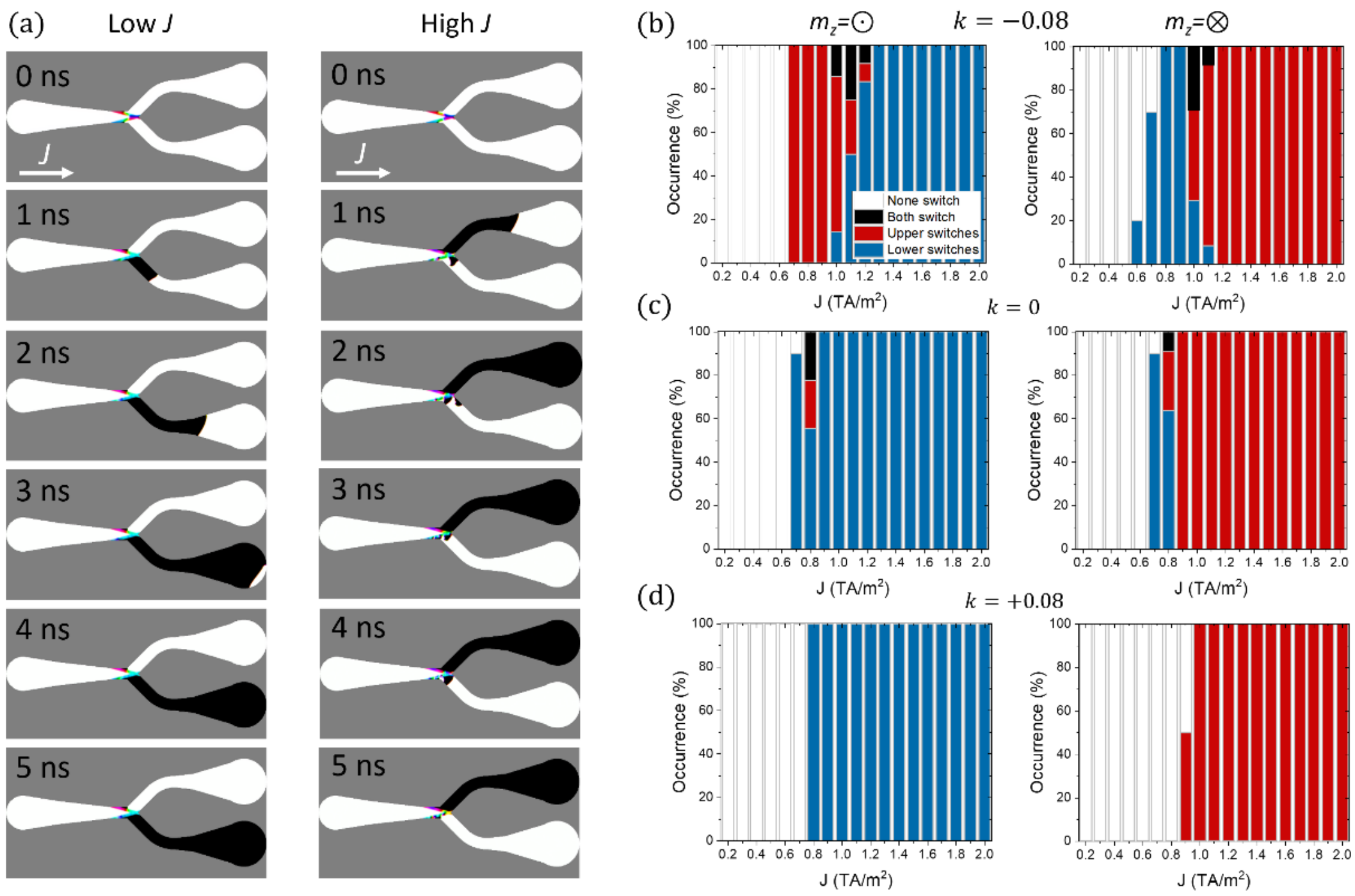

**Fig. 4. Micromagnetic simulations of current-dependent switching in Y-shaped devices.** *(a) Snapshots of the micromagnetic simulations for* $J = 0.7 \times 10^{12}\mathrm{A/m^2}$ *and* $J$=$2.3\times10^{12}\mathrm{A/m^2}$ *for* $k = -0.08$ *after an initialization with +*$H_z$ *and reducing to zero.* (b-d) *Occurrence of magnetization switching as a function of current amplitude for a racetrack initially saturated with +*$H_z$ *(left panel) and -*$H_z$ *(right panel) magnetic field. The data is based on the average of 10 simulations. The ratio between field-like and damping-like torques is (b)* $k = -0.08$*, (c)* $k = 0$ *and (d)* $k = +0.08$.

The fact that we observe single switching in one arm or the other, and no obvious repeated nucleation, calls for a deeper investigation of the physics involved using micromagnetic modeling. The results of the micromagnetic simulations of the Y-shaped devices are summarized in Fig. 4. In particular, at low current densities [left panel of Fig. 4(a)], starting from an ⊙ state, the SOTs trigger inhomogeneous magnetization in the IP region that fluctuates over time [see Fig. 2(e)] in only one (the lower) of the arms while, in the other (upper) arm, the SOTs promote a stable chiral state favored by the DMI. These magnetization fluctuations aid nucleation and the lower arm is the one which switches at low *J*. In the other arm, the SOTs are still too small to promote nucleation at low *J*.

For high *J* [right panel of Fig. 4(a)], the lower arm does not switch. However, magnetization fluctuations are still taking place in the IP region neighboring the lower arm, causing multiple domain nucleation events. Since in curved tracks ⊙|⊗ DWs move faster than ⊗|⊙ DWs [23,42,43], the ⊗ domains are annihilated and so the final state corresponds to the state of the faster DW [22]. The resulting magnetic state in this arm therefore appears as though it has not switched. In the upper arm *J* is now high enough to give SOTs that nucleate a reverse domain.

The asymmetry between ⊙|⊗ and ⊗|⊙ DW velocities can be caused by a number of mechanisms, not just the track curvature. It can also be promoted by the presence of the field-like torque [22], which is often neglected. This can be seen in the simulation results shown in Fig. 4(b-d). If no field-like torque is present [$k = 0$; Fig. 4(c)], the range of current densities where magnetization switching is transferred from one arm to the other is very narrow. If the coefficient $k = +0.08$ [Fig. 4(d)], magnetization switching occurs in only one arm. However, if $k = -0.08$, there is a wider range of current densities where magnetization switching can occur in either arm, which corresponds to our experimental results in Fig. 3. The micromagnetic simulations thus show that the field-like torque also plays an important role in the switching process in addition to the dominating damping-like torque. The strength and sign of the field-like torque is in accordance with the previously measured data [14]. The whole switching mechanism at various current densities can be reversed when starting from the ⊗ state as shown in right-hand panel of Figs. 4(b-d). We have also verified that the effect of the Oersted field on the domain nucleation mechanism is negligible.

In conclusion, we combine experimental data and micromagnetic simulations to show that the SOTs acting on the IP magnetized region in the chirally coupled nanomagnetic system strongly affect the nucleation process in the OOP region, with the type of switching process depending on the geometrical orientation of the IP region. We further show that both the field-like and damping-like SOTs play an important role in the switching process. Such a complex switching dynamics has to be taken into account in devices with embedded chirally-coupled interfaces, since the interaction between the DW and the IP magnetized region will depend on the geometry. Using the Y-shaped devices in conventional Boolean logic [32] requires an equal threshold current in both arms. The independent control of the threshold currents in individual arms can be exploited in non-conventional computing schemes, where the non-equal controlled threshold current for the DW inversion can be utilized for weighted logic [44]. The Y-shaped device also serves as a building block of a magnetic demultiplexer in logic circuits [45], which can distribute a single input to any one of the multiple outputs.

**Acknowledgements**

The authors acknowledge funding from the National Key Research and Development Program of China (No. 2022YFA1203904), the National Natural Science Foundation of China (No. 52271160). The work by E.M. and V.R. was supported by project Projects No. SA114P20 from Junta de Castilla y León, project PID2020117024GB-C41 funded by the Spanish Ministerio de Ciencia e Innovacion (MCIN/AEI/10.13039/501100011033), and project MagnEFi, Grant Agreement No. 860060, (H2020-MSCA-ITN-2019) funded by the European Commission. The data that support this study are available via the Zenodo repository [46].